\documentclass[12pt]{iopart}

\usepackage[T1]{fontenc}
\usepackage{graphicx}
\usepackage{color}
 \expandafter\let\csname equation*\endcsname\relax
 \expandafter\let\csname endequation*\endcsname\relax
\usepackage{amsmath}
\usepackage{bbm}

\input epsf
\begin{document}

\title[Coexistence of spin-triplet superconductivity with magnetism]{Coexistence
of spin-triplet superconductivity with magnetism
within a single mechanism for orbitally degenerate correlated electrons:
Statistically-consistent Gutzwiller
approximation.}

\author{M Zegrodnik$^1$, J Spa\l ek$^{1,2}$, and J B\"{u}nemann$^3$}

\address{$^1$AGH University of Science and Technology,
Faculty of Physics and Applied Computer Science, Al. Mickiewicza 30,
PL-30-059 Krak\'{o}w, Poland}
\address{$^2$Marian Smoluchowski Institute of Physics, 
Jagiellonian University, ul. Reymonta 4,
PL-30-059 Krak\'ow, Poland}
\address{$^3$Max-Planck Institute for Solid State Research,
Heisenbergstr. 1, D-70569 Stuttgart, Germany}

\ead{michal.zegrodnik@gmail.com, ufspalek@if.uj.edu.pl,
buenemann@gmail.com}
\begin{abstract}
An orbitally degenerate two-band Hubbard model is analyzed with inclusion of
the Hund's rule induced
spin-triplet paired states and their coexistence with magnetic ordering. The
so-called \textit{statistically consistent Gutzwiller approximation} (SGA) has
been applied to the case of a square lattice.
The
superconducting gaps, the magnetic moment, and the free energy are analyzed as
a
function of the Hund's rule coupling strength and the band
filling. Also, the
influence
of the
intersite hybridization on the stability of paired phases is discussed. 
In order to examine the effect of
correlations the results are compared with those calculated
earlier
within the Hartree-Fock (HF) approximation combined with the
Bardeen-Cooper-Schrieffer (BCS)
approach.
Significant differences between the two used methods
 (HF+BCS vs. SGA+real-space pairing) appear in the stability regions
of the considered phases. Our results supplement the analysis of this
canonical model used widely in the discussions of pure
magnetic phases with the detailed elaboration of the stability of the
spin-triplet
superconducting states and the
coexistent magnetic-superconducting states. At the end, we
briefly
discuss qualitatively the factors that need to be included for
a detailed quantitative comparison with the corresponding experimental results.
\end{abstract}

\pacs{74.20.-z, 74.25.Dw, 75.10.Lp}
\submitto{\NJP}
\maketitle

\section{Introduction}\label{sec:intro}

The question of coexistence of magnetism and superconductivity
appears very
often in correlated electron systems. In this context, both the spin-singlet
and the spin-triplet paired states should be considered. A general motivation
for considering here the spin-triplet pairing is provided by the
discoveries of superconductivity in
Sr$_2$RuO$_4$ \cite{Maeno1994,Maeno2000}, UGe$_2$ \cite{Saxena2000,Huxley2001},
URhGe \cite{Tateiwa2001}, UIr \cite{Kobayashi2006}, and
UCoGe \cite{Huy2007,Slooten2009,Visser2010}. In
the last four compounds, superconductivity indeed coexists with
ferromagnetism. Moreover, for both, the
spin-singlet high-temperature
superconductors and the heavy-fermion systems, the antiferromagnetism and the
superconductivity can have the same origin. Hence, it is natural to ask whether
 ferromagnetism and spin-triplet
superconductivity also have the same origin in the itinerant uranium
ferromagnets. A related and a very nontrivial question is
concerned with the coexistence of antiferromagnetism with triplet
superconducting state as in UNi$_2$Al$_3$
\cite{Geibel1991,Schroder1994,Ishida2002} and UPt$_3$ \cite{Aeppli1989,Tou1995}.

It has been argued earlier \cite{Klejnberg1999,Spalek2001,Spalek2003,Shen1998}
that for the case of indistinguishable fermions, the
intra-atomic Hund's rule
exchange can lead in a natural manner to
the coexistence of spin-triplet superconductivity with magnetic ordering -
ferromagnetism or antiferromagnetism in the simplest situations. This idea has
been elaborated subsequently by
us \cite{Zegrodnik2011_1,Zegrodnik2011_2,Zegrodnik2012} by means of the combined
Hartree-Fock(HF)-Bardeen-Cooper-Shrieffer(BCS) approach. In particular, the
phase
diagrams have been
determined which contain regions of stability of the pure
superconducting phase of type A (i.e., the equal-spin-paired
phase), as well as
superconductivity coexisting with
either ferromagnetism or antiferromagnetism. 

The HF approximation, as a rule, overestimates the
stability of phases with a
 broken symmetry. Therefore, in this work, we apply the Gutzwiller approximation
for the same selection of phases in order to examine
explicitly
the effects of interelectronic correlations. The
extension of the Gutzwiller
method to the
multi-band case \cite{Bunemann1997,Bunemann1998,Bunemann2005} 
provides us with the so-called renormalization factors for our
degenerate two-band models. With these factors we construct an
effective Hamiltonian by means of
the \textit{statistically consistent Gutzwiller
approximation}, SGA, in which additional constraints are added
to the standard Gutzwiller approximation (GA) and with the incorporation of
which the single-particle state has been determined (see
\cite{Jedrak2010,Kaczmarczyk2011,Howczak2012,Jedrak2011} for
exemplary
applications of the SGA method). The detailed phase diagram and
the corresponding
order
parameters are determined as functions of the microscopic parameters such as the
band
filling, $n$, the Hund's rule exchange integral, $J$, and the Hubbard
interaction parameters, $U$ and $U'$. The obtained results are compared with
those coming out from the
Hartree-Fock approximation. In this manner, the paper extends the discussion of
itinerant magnetism within the canonical (extended Hubbard) model, appropriate
for this purpose, to
the analysis of pure and coexisting superconducting-magnetic states within a
single
unified approach. Additionally, at the end, we dwell briefly on the
applicability of our
original concepts to more realistic systems. It
should be noted that theoretical investigations regarding the spin-triplet
pairing have been performed recently also for other
systems \cite{Sano2003,Han2004,Hotta2004,Dai2008,Lee2008,Imai2012}.

The paper
is composed as follows. In Sec.~II we provide the principal aspects of
real-space spin-triplet pairing induced by the Hund's rule
coupling, and
introduce the band-renormalization factors for our two-band model. 
Furthermore,  in subsections A and B of Sec.~II we 
explain how the effective Hamiltonian is constructed, according
to
the statistically consistent Gutzwiller approximation, for all the phases
considered in this work. In Sec.~III we discuss the phase diagram, and the
principal
order parameters in the considered phases, whereas Sec.~IV contains the
concluding remarks and outlook concerning future investigations to make the
approach applicable to real systems.

\section{Model and method}\label{sec:model}
We consider the extended orbitally-degenerate Hubbard Hamiltonian, which has the
form
\begin{equation}
\begin{split}
\hat{H}&=\sum_{ij(i\neq j)ll^{\prime}\sigma}t^{ll^{\prime}}_{ij}
\hat{c}_{il\sigma}^{\dag}\hat{c}_{jl^{\prime}\sigma}+(U^{\prime}
+J)\sum_{i}\hat{
n } _ { i1}\hat{n}_{
i2}\\
&+U\sum_{il}\hat{n}_{
il\uparrow } \hat { n } _ {
il\downarrow}-J\sum_{ill^{\prime}(l\neq
l^{\prime})}\bigg(\mathbf{\hat{S}}_{il}\cdotp
\mathbf{\hat{S}}_{il^{\prime}}+\frac{3}{4}\hat{n}_{il}\hat{n}_{il^{\prime}}
\bigg)
\label{eq:H_start}\\
&=\hat{H}^0+\hat{H}^{at}\;,
\end{split}
\end{equation}
where $l=1,2$ label the orbitals and the first term describes electron hopping
between atomic sites $i$ and $j$. For $l\neq l^{\prime}$ this term
represents electron hopping with change of the orbital (i.e., 
hybridization in momentum space). The next two terms describe the Coulomb
interactions
between electrons on the same atomic site. However the second
term contains also the contribution, originating from the exchange
interaction
($J$). The last term expresses the Hund's
rule i.e., the ferromagnetic exchange between electrons localized on the same
site, but
on different orbitals. This term contributes to magnetic
coupling and is responsible for the spin-triplet pairing
leading to magnetic ordering, superconductivity and coexistent
magnetic-superconducting phases. 
In the Hamiltonian~(\ref{eq:H_start}), we have disregarded the
pair hopping term 
$(J/2) \sum_{l\neq
l'}\hat{c}^{\dagger}_{il\uparrow}\hat{c}^{\dagger}_{
il\downarrow}\hat{c}_{il'\downarrow}\hat{c}_{il'\uparrow}$ 
 because it hardly influences the ordered phases which we analyze in this
work.
In our variational method we assume that the correlated state $|\Psi_G\rangle$
of the system can be expressed in the following manner
\begin{equation}
|\Psi_G \rangle=\hat{P}_G|\Psi_0 \rangle\;,
\end{equation}
where $|\Psi_0 \rangle$ is the normalized non-correlated state to be determined
later and $\hat{P}_G$
is the
Gutzwiller correlator selected in the following form
\begin{equation}
 \hat{P}_G=\prod_i\hat{P}_{G|i}=\prod_{i}\sum_{I,I^{\prime}}\lambda^{(i)}_{I,I^{
\prime } } |I\rangle_ { ii }
 \langle I^{\prime}|\;.
\label{eq:correlator}
\end{equation}
Here, $\lambda^{(i)}_{I,I^{\prime}}$ are the variational parameters, which are
assumed to be real.
In the two-band situation the local basis consists of 16 states (see Table
\ref{table: values_tab}), which are
defined as follows
\begin{equation}
 |I\rangle_i =\hat{C}^{\dagger}_{i,I}|0\rangle_i \equiv 
\prod_{\gamma \in I}\hat{c}_{i\gamma}^{\dagger}|0\rangle_i=
\hat{c}^{\dagger}_{i\gamma_1}...\hat{c}^{\dagger}_{i{\gamma}_{|I|}
} |0\rangle_i\;,
\label{eq:I_stat}
\end{equation}
where $\gamma=1,2,3,4$ labels the four spin-orbital states
(in the $l\sigma$ notation: $1\uparrow,1\downarrow,2\uparrow,2\downarrow$,
respectively) and $|I|$ is the number of electrons in the local state 
 $|I \rangle$. In
general, an index $I$ can be interpreted as a set in the usual mathematical
sense.
The creation operators in (\ref{eq:I_stat}) are placed in ascending order, i.e.,
$\gamma_1<...<\gamma_{|I|}$. In an analogous
manner, one can define the
product of annihilation operators
\begin{equation}
 \hat{C}_{i,I}=\prod_{\gamma\in
I}\hat{c}_{i\gamma}=\hat{c}_{i\gamma_1}...\hat{c}_{i\gamma_{|I|}}\;,
\end{equation}
which are placed in descending order
$\gamma_1>...>\gamma_{|I|}$. 

\begin{table}[h]
\centering
\caption{The local basis consisting of 16 configurations
containing $N_e=0,...,4$ electrons, which are
enumerated as shown below.}
 \begin{tabular}{||l|l||l|l||l|l||l|l||}
  \hline
       $|0,0\rangle$  &  1     &  $|0,\downarrow\rangle$ & 5
&$|\downarrow,\downarrow\rangle$  &  9     & 
$|\uparrow\downarrow,\uparrow\rangle$ & 13\\ 
       $|\uparrow,0\rangle$  &  2     &  $|\uparrow\downarrow,0\rangle$ & 6
&$|\uparrow,\downarrow\rangle$  &  10     & 
$|\downarrow,\uparrow\downarrow\rangle$ & 14\\ 
       $|0,\uparrow \rangle$  &  3     &  $|0,\uparrow\downarrow\rangle$ & 7
&$|\downarrow,\uparrow\rangle$  &  11    & 
$|\uparrow\downarrow,\downarrow\rangle$ & 15\\ 
       $|\downarrow,0\rangle$  &  4     &  $|\uparrow,\uparrow\rangle$ & 8
&$|\uparrow,\uparrow\downarrow\rangle$  &  12    & 
$|\uparrow\downarrow,\uparrow\downarrow\rangle$ & 16\\ 

  \hline
 \end{tabular}
 \label{table: values_tab}
\end{table}

The operator $|I\rangle_{ii}
\langle I^{\prime}|$ can be expressed in terms of $\hat{C}^{\dagger}_I$ and
$\hat{C}_I$ in the following manner
\begin{equation}
 \hat{m}_{I,I^{\prime}|i}\equiv |I\rangle_{ii}\langle
I^{\prime}|=\hat{C}^{\dagger}_{i,I} \hat{C}_{i,I^{\prime}}\hat{n}^h_{I\cup
I'|i}\;,
\end{equation}
where
\begin{equation}
 \hat{n}^h_{I\cup I'|i}=\prod_{\gamma\in \overline{I\cup
I^{\prime}}}(1-\hat{n}_{i\gamma})\;.
\end{equation}
In the subsequent discussion, we write expectation values with
respect to $|\Psi_0\rangle$ as
\begin{equation}
 \langle\hat{O}\rangle_0=\langle \Psi_0|\hat{O}|\Psi_0\rangle\;,
\end{equation}
while the expectation values with respect to $|\Psi_G\rangle$ will be denoted by
\begin{equation}
 \langle\hat{O}
\rangle_G=\frac{\langle\Psi_G|\hat{O}|\Psi_G\rangle}{\langle\Psi_G|\Psi_G\rangle
}\;.
\end{equation}
The most important step within the Gutzwiller approach is to
derive the formula
for the expectation value of the Hamiltonian $\hat{K}=\hat{H}-\mu \hat{N}$ with
respect to $|\Psi_G\rangle$. 
This can be done in the limit of infinite dimensions by a 
diagrammatic approach \cite{Bunemann2005} which uses the variational analog of
Feynmann diagrams.
By applying this method to the interaction part of the Hamiltonian
(\ref{eq:H_start}), which is completely of intra-site
character, one obtains
\begin{equation}
\langle \hat{H}^{at} \rangle_{G}=L\sum_{I_1,I_4}\bar{E}_{I_1,I_4}
\langle\hat{m}_{I_1,I_4}
\rangle_0\;,
\label{eq:H_atomic}
\end{equation}
where
\begin{equation}
 \bar{E}_{I_1,I_4}=\sum_{I_2,I_3}\lambda_{I_1,I_2}\lambda_{I_3,I_4}\langle
I_2|\hat{H}^{at}|I_3\rangle\;,
\end{equation}
and $L$ is the number of atomic sites. In (\ref{eq:H_atomic}) we have
assumed that our system is homogeneous. Note that,   
with the use of Wick's theorem, the purely local expectation values
$\langle\hat{m}_{I_1,I_4}\rangle_0$ can be expressed
in terms of the local single-particle density matrix elements
$\langle\hat{c}^{\alpha}_{i\gamma}
\hat{c}^{\alpha'}_{i\gamma'}\rangle_0$.  Here, 
$\hat{c}^{\alpha}_{i\gamma}$ are either creation or annihilation 
operators.

The expectation value of the single-particle part in the 
 Hamiltonian (\ref{eq:H_start}) can be cast to the form
\begin{equation}
 \langle \hat{H^0}\rangle_G=\sum_{ij(i\neq
j)}\sum_
{\gamma
\gamma^{\prime}\tilde{\gamma}\tilde{\gamma}'}t^{\gamma \gamma^{\prime}}_{ij}
\big(q_{\gamma \tilde{\gamma}}q_{\gamma' \tilde{\gamma}'}
-\bar{q}_{\gamma \tilde{\gamma}}\bar{q}_{\gamma' \tilde{\gamma}'}\big) \langle
\hat{c}^{\dagger}_{i,\tilde{\gamma}}
\hat{c}_{j,\tilde{\gamma}^{\prime}} \rangle_0
\label{eq:el_hopping}
\end{equation}
where we have assumed that the renormalization factors
$q$ and $\bar{q}$ are real
numbers and
$t^{\gamma \gamma'}=t^{\gamma'\gamma}$. Moreover,
in the equation above we have neglected the part containing
the inter-site pairing terms
$\langle \hat{c}^{\dagger}_{i,\gamma}\hat{c}^{\dagger}_{j,\gamma^{\prime}
} \rangle_0$ and
$\langle \hat{c}_{i,\gamma}\hat{c}_{j,\gamma^{\prime}
} \rangle_0$ as we are going to concentrate on the Hund's rule induced
intra-site spin-triplet paired states. The inter-site pairing
amplitudes are much smaller than the intra-site terms, 
 in the considered model. The renormalization factors, introduced in (\ref{eq:el_hopping}), have the form
\begin{equation}
 q_{\gamma\tilde{\gamma}}=\sum_{I(\tilde{\gamma} \notin
I)}\bigg[\sum_{I'}\mbox{fsgn}(\tilde{\gamma},I)m^{
0(\tilde{\gamma}) }_ {I,I'}c^*_{I\cup\tilde{\gamma},I'|\gamma}+\sum_{I'(\tilde{\gamma}
\notin
I')}\mbox{fsgn}(\tilde{\gamma},I)m^{0}_{I',I\cup\tilde{\gamma}}
c^*_{I',I|\gamma}\bigg]\;,
\label{eq:q_ren_fac}
\end{equation}
where $m^0_{I,I'}=\langle \hat{m}_{I,I'} \rangle_0$ and
$m^{0(\tilde{\gamma})}_{I,I'}=\langle \hat{m}^{(\tilde{\gamma})}_{I,I'} \rangle_0$. Here we have
introduced the operator
\begin{equation}
 \hat{m}^{(\gamma)}_{I,I'}=\hat{C}^{\dagger}_{i,I}
\hat{C}_{i,I^{\prime}}\hat{n}^h_{I\cup
I'\cup\gamma|i}\;.
\end{equation}
The parameters $c^*_{I_1,I_2|\gamma}$ in (\ref{eq:q_ren_fac}) are defined as
\begin{equation}
 c^*_{I_1,I_2|\gamma}=\sum_{I(\gamma\notin
I)}\mbox{fsgn}(\gamma,I)\lambda_{I_1,I\cup\gamma}\lambda_{I,I_2}\;
,
\end{equation}
where we introduced the fermionic sign function
\begin{equation}
\mbox{fsgn}(\gamma,I)\equiv  \langle  I\cup \gamma |
 \hat{c}^{\dagger}_{\gamma}|I\rangle\;.
\end{equation}
The renormalization factors $\bar{q}_{\gamma \tilde{\gamma}}$ have to be
included in (\ref{eq:el_hopping}), when there are nonzero gap parameters
 ($\langle \hat{c}^{\alpha}  \hat{c}^{\alpha} \rangle_{0}\neq 0  $)
 in
$|\Psi_0\rangle$, which is the case considered here. The form of
$\bar{q}_{\gamma \tilde{\gamma}}$ is as follows
\begin{equation}
 \bar{q}_{\gamma \tilde{\gamma}}=\sum_{I(\tilde{\gamma} \notin
I)}\bigg[\sum_{I'}\mbox{fsgn}(\tilde{\gamma},I)m^{
0(\tilde{\gamma})}_ {I',I}c^*_{I',I\cup\tilde{\gamma}|\gamma}
+\sum_{I'(\tilde{\gamma} \notin
I')}\mbox{fsgn}(\tilde{\gamma},I)m^{0}_{I\cup\tilde{\gamma},I'}
c^*_{I,I'|\gamma} \bigg ]\;.
\label{eq:qbar_ren_fac}
\end{equation}
The remaining part of $\langle \hat{K} \rangle_G$ that has to be derived is
the expectation value $\langle \hat{N}\rangle_G$. Also in this case, the
diagrammatic evaluation in infinite dimensions gives the proper formula, 
\begin{equation}
 \langle \hat{N}\rangle_G=\sum_{i\gamma}\langle
\hat{n}_{i\gamma}
\rangle_G\;,
\label{eq:N_G}
\end{equation}
where
\begin{equation}
\langle
\hat{n}_{i\gamma}
\rangle_G=\sum_{I_1,I_4}N^{\gamma}_{I_1,I_4}m^0_{
I_1,I_4
}\;,
\label{eq:N_G_2}
\end{equation}
and
\begin{equation}
 N^{\gamma}_{I_1,I_4}=\sum_{I(
\gamma\notin I)}\lambda_{I_1,I\cup \gamma}\lambda_{I\cup
\gamma,I_4}\;.
\end{equation}
The pairing densities in the correlated state that are going to be useful in the
subsequent discussion can be expressed in the following way
\begin{equation}
 \langle\hat{c}_{i\gamma}\hat{c}_{i\gamma'}
\rangle_G=\sum_{I_1,I_4}S^{\gamma \gamma'}_{I_1,I_4}m^0_{I_1,I_4}\;,
\label{eq: delt_G}
\end{equation}
where
\begin{equation}
 S^{\gamma \gamma'}_{I_1,I_4}=\sum_{I(\gamma \gamma'\notin I)}\lambda_{I_1,I}
\lambda_{I\cup(\gamma\gamma'),I_4}\mbox{fsgn}(\gamma,I)\mbox{fsgn}
(\gamma',I)\mbox{fsgn}(\gamma',\gamma)\;.
\end{equation}

Using (\ref{eq:H_atomic}), (\ref{eq:el_hopping}), and (\ref{eq:N_G})
one can express $\langle \hat{K} \rangle_G$ in terms of the variational
parameters
$\lambda_{I,I'}$, local and non-local single
particle density matrix elements ,$\langle\hat{c}^{\alpha}_{i\gamma}
\hat{c}^{\alpha'}_{i\gamma'}\rangle_0$,
$\langle \hat{c}^{\dagger}_{i,\gamma} \hat{c}_{j,\gamma^{\prime}} \rangle_0$,
and the matrix elements of the atomic part of the atomic
Hamiltonian
represented in the local basis $\langle I|\hat{H}^{at}|I'\rangle$.

The formula for $\langle \hat{K}\rangle_G$, obtained in the way
described above, can be written as an expectation value of an effective
Hamiltonian $\hat{K}_{GA}$, evaluated with respect to
$|\Psi_0\rangle$
\begin{equation}
\begin{split}
 \hat{K}_{GA}&=
\sum_{ij(i\neq
j)}\sum_
{\gamma
\gamma^{\prime}\tilde{\gamma}\tilde{\gamma}'}t^{\gamma \gamma^{\prime}}_{ij}
\big(q_{\gamma \tilde{\gamma}}q_{\gamma' \tilde{\gamma}'}
-\bar{q}_{\gamma \tilde{\gamma}}\bar{q}_{\gamma' \tilde{\gamma}'}\big)
\hat{c}^{\dagger}_{i,\tilde{\gamma}}
\hat{c}_{j,\tilde{\gamma}^{\prime}}
\\
&\quad\quad-\mu
\sum_{i\gamma}q^s_{\gamma}\hat{n}_{i\gamma}
+L\sum_{I_1,I_4} \bar{E}_{I_1,I_4}
\langle\hat{m}_{I_1,I_4}
\rangle_0,
\label{eq:U_GA}
\end{split}
\end{equation}
where $q^s_{\gamma}=\langle \hat{n}_{i\gamma}
\rangle_G\slash \langle \hat{n}_{i\gamma} \rangle_0$.
There is no guarantee that the condition
\begin{equation}
 \langle \hat{n}_{i\gamma}\rangle_G= \langle \hat{n}_{i\gamma}
\rangle_0\;,
\end{equation}
is fulfilled. It turns out that it is fulfilled for the paramagnetic and the
magnetically ordered phases of our two-band system, however it is not for the
superconducting phases. Physically it is most sensible to fix
$\langle\hat{n}\rangle_G$ instead of $\langle\hat{n} \rangle_0$, during the
minimization. This is the reason why we
include the term $-\mu\hat{N}$ already at the beginning of our derivation in
$\langle \hat{K} \rangle_G$. In this manner the chemical
potential $\mu$ refers to the
initial correlated system, not to the effective non-correlated one (for which
the chemical potential can be different).

Having in mind that there are 16 states in the local basis there could be up to
$16\times 16=256$ variational parameters $\lambda_{I,I'}$. However, for
symmetry reasons many of these parameters are zero. The finite
parameters can be
identified by the following rule
\begin{equation}
 \lambda_{I,I^{\prime}}\neq 0  \Leftrightarrow \langle \hat{m}_{I,I'}\rangle_0
\neq 0 \vee \langle I|\hat{H}^{at}|I'\rangle \neq 0;.
\label{eq:assum_lambda}
\end{equation}
It should also be noted that, as shown in \cite{Bunemann2005}, the
variational                                   
parameters are not
independent since they have to obey the constrains
\begin{equation}
\begin{split}
 \langle\hat{P}^2_{G|i}\rangle_0&=1,\\
 \langle\hat{c}^{\dagger}_{i\gamma}\hat{P}^2_{G|i}\hat{c}_{i\gamma'
}\rangle_0&=\langle\hat{c}^{\dagger}_{i\gamma}\hat{c}_{i\gamma'}
\rangle_0,\\
 \langle\hat{c}^{\dagger}_{i\gamma}\hat{P}^2_{G|i}\hat{c}^{\dagger}_{
i\gamma'}\rangle_0&=\langle\hat{c}^{\dagger}_{i\gamma}\hat{c}^{\dagger}_{
i\gamma' }
\rangle_0,\\
 \langle\hat{c}_{i\gamma}\hat{P}^2_{G|i}\hat{c}_{
i\gamma'}\rangle_0&=\langle\hat{c}_{i\gamma}\hat{c}_{
i\gamma' }
\rangle_0\;,\\
\end{split}
\end{equation}
which are going to be used to fix some of the parameters
$\lambda_{I,I'}$.

The results presented in this work have been obtained for the case of a
square lattice with the band dispersions
\begin{equation}
 \epsilon_{1\mathbf{k}}=\epsilon_{2\mathbf{k}}\equiv
\epsilon_{\mathbf{k}}=2t(\cos{(k_x)}+\cos{(k_y)})\;,
\label{eq:deg_band}
\end{equation}
and also
\begin{equation}
 \epsilon_{12\mathbf{k}}=\epsilon_{21\mathbf{k}}=\beta_h\epsilon_\mathbf{k}\;,
\label{eq:hyb_beta}
\end{equation}
where $\beta_h\in [0,1]$. The orbital degeneracy and spatial homogeneity
allow us to write
\begin{equation}
\begin{split}
\langle \hat{n}_{i1} \rangle_G&=\langle \hat{n}_{i2} \rangle_G\equiv n_G/2, \\
\langle \hat{S}^z_{i1} \rangle_G&=\langle \hat{S}^z_{i2} \rangle_G\equiv
S^z_G\;,
\end{split}
\label{eq:nn_G}
\end{equation}
where
\begin{equation}
\begin{split}
\hat{S}^z_{il}&\equiv\frac{1}{2}\big(
\hat{n}_{il\uparrow}-\hat{n}_{il\downarrow}
\big),\\ 
\hat{n}_{il}&\equiv\hat{n}_{il\uparrow}+\hat{n}_{il\downarrow}\;.
\label{eq: S_def}
\end{split}
\end{equation}
Similar expressions as in (\ref{eq:nn_G}) can be introduced for the expectation
values in the non-correlated state $|\Psi_0\rangle$.

Before discussing the principal magnetic and/or spin-triplet superconducting
phases, we introduce first the exact expression of the full exchange operator
(the last term of our Hamiltonian) via the local spin-triplet pairing operators
($\hat{A}^{\dagger}_{im}$, $\hat{A}_{im}$) namely
\begin{equation}
 \sum_{ll^{\prime}(l\neq
l^{\prime})}\bigg(\mathbf{\hat{S}}_{il}\cdotp
\mathbf{\hat{S}}_{il^{\prime}}+\frac{3}{4}\hat{n}_{il}\hat{n}_{il^{\prime}}
\bigg)=\sum_m\hat{A}^{\dagger}_{im}\hat{A}_{im}\;,
\end{equation}
where
\begin{equation}
\hat{A}^{\dagger}_{i,m}\equiv\left\{\begin{array}{cl}
a^{\dagger}_{i1\uparrow}a^{\dagger}_{i2\uparrow} & m=1\\
a^{\dagger}_{i1\downarrow}a^{\dagger}_{i2\downarrow} & m=-1\\
\frac{1}{\sqrt{2}}(a^{\dagger}_{i1\uparrow}a^{\dagger}_{i2\downarrow}+a^{\dagger
}_{i1\downarrow}a^{\dagger}_{i2\uparrow}) & m=0\;.\\
\end{array}\right.
\label{eq:paring_op}
\end{equation}
We see that those two representations are mathematically equivalent, so the
phase with $S^z_G=\langle\hat{S}_{il}^z\rangle_G \neq 0$ and that with the
corresponding off-diagonal order parameter $\langle
\hat{A}_{im} \rangle_G \neq 0$
(or $\langle\hat{A}^{\dagger}_{im}
\rangle_G \neq 0$) should be treated on equal footing.
%%%%%%%%%%%%%%%%%%%%%%%%%%%%%%%%%%%%%%%%%%%%%%%%%%%%%%%%%%%%%%%%%%%%%%%%%%%%%%%%
\subsection{Statistically-consistent Gutzwiller method for superconducting and
coexistent
superconducting-ferromagnetic phases}

In this subsection we will describe the SGA approach as applied to the selected phases characterized by the following order
parameters
\begin{itemize}
\item  Superconducting phase of type A1 coexisting with ferromagnetism
(A1+FM):\newline
$S_{G|u}^{z}\neq 0$, $\Delta^{G}_1\neq 0$, $\Delta^{G}_{-1}=\Delta^{G}_0=0$,
\item  Pure type A superconducting phase (A):\newline
$S_{G|u}^{z}=0$, $\Delta^{G}_1=\Delta^{G}_{-1}\neq 0$, $\Delta^{G}_0=0$,
\item  Pure ferromagnetic phase (FM):\newline
$S_{G|u}^{z}\neq 0$, $\Delta^{G}_1=\Delta^{G}_{-1}=\Delta^{G}_0=0$,
\item  Paramagnetic phase (NS): \newline
$S_{G|u}^{z}= 0$, $\Delta^{G}_1=\Delta^{G}_{-1}=\Delta^{G}_0=0$,
\end{itemize}
where $S_{G|u}^z$ refers to the uniform magnetic moment and 
\begin{equation}
 \Delta_{m}^G=\langle\hat{A}_{im} \rangle_G, \quad
(\Delta^{G}_{m})^{*}=\langle\hat{A}^{\dagger}_{im}
\rangle_G\;,
\end{equation}
are the spin-triplet local gap parameters which are assumed as
real here.

The (correlated) order parameters which have been used above 
to define the relevant phases can also be  defined for the 
non-correlated state $|\Psi_0\rangle$. With these, we can determine which of
the matrix elements $\langle \hat{m}_{I,I'} \rangle_0$ are equal to zero for
the considered phases. The assumption (\ref{eq:assum_lambda})
then allows us to choose the non-diagonal variational parameters,
$\lambda_{I,I'}$, that
have to be taken into account during the calculations. We list
their indexes $(I,I')$ in Table \ref{table: var_parqam}.
\begin{table}[h]
\centering
\caption{Nonzero, off-diagonal local variational parameters
($\lambda_{I,I'}=\lambda_{I',I}$) that are used in the calculations for
the considered phases.}
 \begin{tabular}{||c||c|c|c|c|c|c|c|c|c|c|c||}
  \hline
   $I $ &1 &2 &3 &4 &5 &8 &9 &8 &10 & 1& 1 \\
  \hline
   $ I' $ &16 &15 &14 &13 &12 &16 &16 &9 &11 & 8& 9 \\
  \hline
 \end{tabular}
 \label{table: var_parqam}
\end{table}

As one can see from Table \ref{table: var_parqam}, the off-diagonal variational
parameters correspond to the
creation or annihilation of the Cooper pair in the proper
spin-triplet states $|1\uparrow, 2\uparrow\rangle$ and $|1\downarrow,
2\downarrow\rangle$  (phase A). Because in the A1 phase only electrons with
spin-up are paired one can assume that $\lambda_{1,16}$, $\lambda_{2,15}$,
$\lambda_{3,14}$, $\lambda_{8,16}$, $\lambda_{8,9}$ are zero
(and their transoposed corespondants - $\lambda_{I,I'}=\lambda_{I',I}$). For the
FM and NS unpaired states only
$\lambda_{10,11}$ and $\lambda_{11,10}$ are nonzero. They correspond to the
two
non-diagonal matrix elements of the atomic Hamiltonian, 
 $\langle
I|\hat{H}^{at}|I'\rangle $. With the information contained in Table
\ref{table:
var_parqam}, one obtains the following relations regarding the band-narrowing
renormalization factors
\begin{equation}
\begin{split}
 q_{l\sigma,l'\sigma'}\neq 0 \Leftrightarrow l=l'\wedge
\sigma=\sigma',\\
 \bar{q}_{l\sigma,l'\sigma'}\neq 0 \Leftrightarrow l\neq l'\wedge
\sigma=\sigma'\;,\\
\end{split}
\end{equation}
where we have again used the $\gamma=l\sigma$ notation.
Due to the degeneracy of our bands we find
\begin{equation}
\begin{split}
 &q_{1\sigma,1\sigma}=q_{2\sigma,2\sigma}\equiv q_{\sigma}\;,\\
 &\bar{q}_{2\sigma,1\sigma}=\bar{q}_{1\sigma,2\sigma}\equiv
\bar{q}_{\sigma}\;,\\
&q^s_{1\sigma}=q^s_{2\sigma}\equiv q^s_{\sigma}\;.
\end{split}
\label{eq:q_param_as}
\end{equation}
Using the equations above we can rewrite the Hamiltonian
(\ref{eq:U_GA}) in the more explicit form, in reciprocal space
\begin{equation}
%\begin{split}
 \hat{K}_{GA}=\sum_{\mathbf{k}l\sigma}(Q_{\sigma}\epsilon_{\mathbf{k}}-q^s_{
\sigma}\mu)
\hat{n}_{\mathbf{k}l\sigma}
+\sum_{\mathbf{k}ll'\sigma}Q_{\sigma}\epsilon_{\mathbf{k}12}
\hat{c}^{\dagger}_{\mathbf{k}l\sigma}\hat{c}_{\mathbf{k}l'\sigma}+L\sum_{I_1,I_4
}
\bar{E}_{I_1,I_4}
\langle\hat{m}_{I_1,I_4}
\rangle_0\;,
\label{eq:U_GA_recip}
%\end{split}
\end{equation}
where the renormalization factors $Q_{\sigma}$ are defined as
\begin{equation}
 Q_{\sigma}\equiv q^2_{\sigma}-\bar{q}^2_{\sigma}\;.
\end{equation}

Having the formula for $\hat{K}_{GA}$, given by (\ref{eq:U_GA_recip}), one can
introduce next the so-called
\textit{statistically-consistent Gutzwiller
approximation} (SGA). In this method, the mean fields (such as the expectation
values
for magnetization or superconducting gaps) are treated as variational
mean-field order parameters
with respect to which the energy of the system is minimized. However, in order
to make sure that they coincide with the corresponding values
calculated self-consistently, additional constraints have to be introduced with
the
help of the Lagrange-multiplier
method \cite{Jedrak2010,Kaczmarczyk2011,Howczak2012,Jedrak2011}.
This leads to supplementary terms in the
effective
Hamiltonian of the following form
\begin{equation}
\begin{split}
 \hat{K}_{\lambda}=\hat{K}_{GA} &- \sum_{m=\pm
1}\bigg[\lambda_m\bigg(\sum_\mathbf{k}
\hat{A}_{\mathbf{k}m} -
L\Delta^0_m\bigg)+H.C.\bigg] \\
&-\lambda_S\bigg(\sum_{\mathbf{k}l}\hat{S}^z_{\mathbf{k}l}-2LS_0^z\bigg)
-\lambda_n\bigg(\sum_{\mathbf{k}l\sigma}q^s_{l\sigma}\hat{n}_{
kl\sigma}
-Ln_G\bigg)\;,
\end{split}
\end{equation}
where the Lagrange multipliers $\lambda_m$, $\lambda_s$, and $\lambda_n$ are
introduced to assure that the averages $\langle\hat{A}_{\mathbf{k}m}\rangle$,
$\langle\hat{S}_{\mathbf{k}l}\rangle$ and
$\langle\hat{n}_{\mathbf{k}l\sigma}\rangle$ calculated either from the
corresponding
self-consistent equations or variationally, coincide with each other
\cite{Jedrak2011}.

Introducing the four-component representation of single-particle operators
\begin{equation}
 \mathbf{\hat{f}}^{\dagger}_{\mathbf{k}\sigma}=(\hat{c}^{\dagger}_{\mathbf{k}1
\sigma } ,
 \hat{c}^{\dagger}_{\mathbf{k}2\sigma}, \hat{c}_{-\mathbf{k}1\sigma},
 \hat{c}_{-\mathbf{k}2\sigma})\;,
\end{equation}
we can write down the effective Hamiltonian in the following form
\begin{equation}
\begin{split}
 \hat{K}_{\lambda}&=\frac{1}{2}\sum_{\mathbf{k}\sigma}\mathbf{\hat{f}}^{\dagger}
_{\mathbf{k } \sigma }
\mathbf{\hat{M}}_{\mathbf{k}\sigma}\mathbf{\hat{f}}_{\mathbf{k}\sigma}
+\sum_{\mathbf{k}\sigma} \tilde{\epsilon}_{\mathbf{k}\sigma} 
+2L\sum_{m=\pm1}\lambda_m\Delta_m^0 + 2L\lambda_S S^z_0 + L\lambda_n n_G\\
&+L\sum_{I_1,I_4}
\bar{E}_{I_1,I_4}
\langle\hat{m}_{I_1,I_4}
\rangle_0\;,
\end{split}
\label{eq:eff_Ham_1}
\end{equation}
where $\mathbf{\hat{M}}_{\mathbf{k}\sigma}$ is a 4x4 orthogonal matrix
\begin{equation}
\mathbf{\hat{M}}_{\mathbf{k}\sigma}=\left(\begin{array}{cccc}
\tilde{\epsilon}_{\mathbf{k}\sigma} &
Q_{\sigma}\epsilon_{\mathbf{k}12} & 0 &
\lambda_{\sigma}\\
Q_{\sigma}\epsilon_{\mathbf{k}12} &
\tilde{\epsilon}_{\mathbf{k}\sigma} & -\lambda_{\sigma} & 0\\
0 & -\lambda_{\sigma} & -\tilde{\epsilon}_{\mathbf{k}\sigma} &
-Q_{\sigma}\epsilon_{\mathbf{k}12}\\
\lambda_{\sigma} & 0 &
-Q_{\sigma}\epsilon_{\mathbf{k}12}
&-\tilde{\epsilon}_{\mathbf{k}\sigma}
\end{array} \right)\;.
\label{eq:matrix_H}
\end{equation}
Here we introduced $\lambda_{\uparrow}$ and
$\lambda_{\downarrow}$ 
 which correspond
to the Lagrange parameters $\lambda_{m=1}$ and
$\lambda_{m=-1}$, respectively.
The bare quasiparticle energies $\tilde{\epsilon}_{\mathbf{k}l\sigma}$
 are
defined as 
\begin{equation}
 \tilde{\epsilon}_{\mathbf{k}\sigma}=Q_{\sigma}\epsilon_{\mathbf{k}}-q_
{\sigma} ^s(\mu+\lambda_n)-\frac{1}{2}\sigma\lambda_S\;.
\end{equation}
The diagonalization of the matrix (\ref{eq:matrix_H}) yields
the quasiparticle
eigen-energies in the paired states of the following form 
\begin{equation}
\begin{split}
 E_{\mathbf{k}1\sigma}&=\sqrt{\tilde{\epsilon}_{\mathbf{k}\sigma}^2+\lambda_{
\sigma}^2 } -Q_ { \sigma }\epsilon_{\mathbf{k}12}\;,\\
 E_{\mathbf{k}2\sigma}&=\sqrt{\tilde{\epsilon}_{\mathbf{k}\sigma}^2+\lambda_{
\sigma}^2 } +Q_ { \sigma }\epsilon_{\mathbf{k}12}\;,\\
 E_{\mathbf{k}3\sigma}&=-\sqrt{\tilde{\epsilon}_{\mathbf{k}\sigma}^2+\lambda_{
\sigma}^2 } -Q_ { \sigma }\epsilon_{\mathbf{k}12}\;,\\
 E_{\mathbf{k}4\sigma}&=-\sqrt{\tilde{\epsilon}_{\mathbf{k}\sigma}^2+\lambda_{
\sigma}^2 } +Q_ { \sigma }\epsilon_{\mathbf{k}12}\;.\\
\end{split}
\end{equation}
The first two energies correspond to the doubly degenerate spin-split
quasiparticle excitations in the A phase, 
whereas the remaining two are their
quasihole correspondents.

Even though the Gutzwiller approach was derived for zero temperature, we may
still construct the
grand-potential function $F_{\lambda}$ (per atomic site) that corresponds to the
effective Hamiltonian (\ref{eq:eff_Ham_1}), i.e.,
\begin{equation}
\begin{split}
 F_{\lambda}&=-\frac{1}{L\beta}\sum_{\mathbf{k}l\sigma}\ln\big[1+e^{-\beta
E_{\mathbf{k}l\sigma}} \big]+\frac{1}{L}\sum_{\mathbf{k}\sigma}
\tilde{\epsilon}_{\mathbf{k}\sigma}+2\sum_{m=\pm1}\lambda_m\Delta_m^0 +
2\lambda_S S^z_0 + (\lambda_n+\mu) n_G\\
&+\sum_{I_1,I_4}
\bar{E}_{I_1,I_4}
\langle\hat{m}_{I_1,I_4}
\rangle_0\;.
\end{split}
\end{equation}
The values of the mean fields, the variational parameters, and the Lagrange
multipliers
are found by minimizing the $F_{\lambda}$ functional, i.e., the necessary
conditions for minimum are
\begin{equation}
 \frac{\partial F_{\lambda}}{\partial \vec{A}}=0\;, \quad \frac{\partial
F_{\lambda}}{\partial \vec{\Lambda}_V}=0\;, \quad \frac{\partial
F_{\lambda}}{\partial \vec{\Lambda}_L}=0\;,
\label{eq:derivs}
\end{equation}
where $\vec{A}$, $\vec{\Lambda}_V$, $\vec{\Lambda}_L$ denote collectively the
mean fields in the non-correlated state, the variational parameters and the
Lagrange multipliers respectively. Additionally, the chemical potential, $\mu$
enters through the relation
\begin{equation}
 \frac{\partial F_{\lambda}}{\partial n_G}=\mu\;.
\label{eq:derivs_mu}
\end{equation}
After
solving the complete set of equations, one still has to calculate the mean fields
in the
correlated state with the use of their analogs in the non-correlated state and
the variational parameters using (\ref{eq:N_G_2}) and (\ref{eq:
delt_G}).

With the SGA method one minimises the variational ground 
state energy $\langle \hat{K}  \rangle_{G}$ with respect to the variational 
parameters $\lambda_{I,I'}$ and the single-particle states  $|\Psi_0\rangle$.
Note that an alternative way for this minimization has been introduced, 
e.g., in \cite{buenemann2010}. 
Beyond the ground-state properties of $\hat{K}$ one is often 
 also interested in the (effective) single-particle Hamiltonian 
(\ref{eq:eff_Ham_1}) because its eigenvalues 
are  interpreted as quasi-particle 
excitation energies \cite{thul2003}.

%%%%%%%%%%%%%%%%%%%%%%%%%%%%%%%%%%%%%%%%%%%%%%%%%%%%%%%%%%%%%%%%%%%%%%%%%%%%%%%%
\subsection{Statistically-consistent Gutzwiller method for the coexistent
antiferromagnetic-spin-triplet
superconducting phase}

To consider antiferromagnetism in the simplest case, we 
divide our system
into two interpenetrating sublattices A and B. In
accordance with this division, we define the annihilation
operators on the
sublattices
\begin{equation}
\hat{c}_{il\sigma}=\left\{\begin{array}{cl}
\hat{c}_{il\sigma A} \quad \textrm{for i} \in A\;,\\
\hat{c}_{il\sigma B} \quad \textrm{for i} \in B\;.\\
\end{array}\right.
\end{equation}
The same holds for the creation operators. Next, the Gutzwiller correlator
can be expressed in the form
\begin{equation}
 \hat{P}_{G}=\prod_{i(A)}\hat{P}_{G|i}^{(A)}\prod_{i(B)}\hat{P}_{G|i}^{(B)}\;,
\end{equation}
where
\begin{equation}
 \hat{P}_{G|i}^{(A/B)}=\sum_{I,I'}\lambda^{(A/B)}_{I,I'}|I\rangle_{ii}
\langle I'|\;.
\label{eq:correl_anti}
\end{equation}
If we assume that charge ordering is absent, we have
\begin{equation}
 \langle \hat{S}^z_{ilA}\rangle_G \equiv S^z_{G|s}, \quad
 \langle \hat{S}^z_{ilB}\rangle_G \equiv -S^z_{G|s}\;,
\end{equation}
\begin{equation}
 \langle \hat{n}_{ilA}\rangle_G=\langle n_{ilB}\rangle_G\equiv n_G/2\;. 
\end{equation}
Similar expressions can be obtained for the case of expectation
values taken
in the state $|\Psi_0 \rangle$.
As one can see from (\ref{eq:correl_anti}), we have introduced separate sets of
variational parameters
($\lambda_{I,I'}^A$ and $\lambda_{I,I'}^B$) for the two sublattices.
Fortunately, it does not mean that we have twice as many variational parameters
as in the preceding subsection. The parameters $\lambda_{I,I'}^A$ are related
to the corresponding $\lambda_{I,I'}^B$ through  
\begin{equation}
 \lambda_{I_1,I_2}^{(A)}=\lambda_{I_3,I_4}^{(B)}\;,
\end{equation}
where the states $I_1$ and $I_2$ have opposite spins to those in
the $I_3$ and $I_4$ states, respectively. The same division has to be made for
the renormalization factors $q$, $\bar{q}$ and $q^s$.  They
fulfill the transformation relations
\begin{equation}
\begin{split}
 q^A_{\gamma,\gamma'}=q^B_{\bar{\gamma},\bar{\gamma}'},\\
 \bar{q}^A_{\gamma,\gamma'}=\bar{q}^B_{\bar{\gamma},\bar{\gamma}'},\\
 q^{s}_{\gamma A}=q^{s}_{\bar{\gamma} B}\;,
\end{split}
\end{equation}
where $\gamma$ and $\bar{\gamma}$  are spin-orbitals with
opposite spins. The coexistent
superconducting-antiferromagnetic phase (SC+AF) can be defined in the following
way
\begin{equation}
\begin{split}
 \Delta^G_{1A}=\Delta^G_{-1B}&\equiv \Delta^G_{+}\neq 0\;,\\ 
 \Delta^G_{-1A}=\Delta^G_{1B}&\equiv \Delta^G_{-}\neq 0\;,\\
  S^z_{G|s}&\neq 0\;.
\end{split}
\end{equation}
Considerations analogical to those presented in subsection 2 lead to the
conclusion that for both sublattices the non-diagonal variational parameters,
$\lambda^A_{I,I'}$ and $\lambda^B_{I,I'}$, that have to be used in the
calculations, appropriate for the SC+AF phase, are 
the same as those listed in
Table \ref{table:
var_parqam}. This fact, and the degeneracy of our bands,
 allow us to apply (\ref{eq:q_param_as}) for both
sets of renormalization factors (for A and B sublattices), as we have
\begin{equation}
\begin{split}
 &q^A_{1\sigma,1\sigma}=q^A_{2\sigma,2\sigma}=q^B_{1\bar{\sigma},1\bar{\sigma}}
=q^B_ { 2\bar{\sigma} ,2\bar{\sigma}}\equiv q_{\sigma}\;,\\
 &\bar{q}^A_{2\sigma,1\sigma}=\bar{q}^A_{1\sigma,2\sigma}=\bar{q}^B_{2\bar{
\sigma} , 1\bar{\sigma} } =\bar{q}^B_{1\bar{\sigma},2\bar{\sigma}}\equiv
\bar{q}_{\sigma}\;,\\
&q^s_{1\sigma A}=q^s_{2\sigma A}=q^s_{1\bar{\sigma} B}=q^s_{2\bar{\sigma}
B}\equiv q^s_{\sigma}\;,
\end{split}
\label{eq:q_param_as2}
\end{equation}
where $\bar{\sigma}$ represents the spin opposite to $\sigma$.
Now, we can write down the Hamiltonian $\hat{K}_{GA}$ for the case of SC+AF
phase
\begin{equation}
\begin{split}
 \hat{K}_{GA}&=\sum_{\mathbf{k}l\sigma}Q\epsilon_{\mathbf{k}}
(\hat{c}^{\dagger}_{\mathbf{k}l\sigma A}\hat{c}_{\mathbf{k}l\sigma
B}+\hat{c}^{\dagger}_{\mathbf{k}l\sigma
B}\hat{c}_{\mathbf{k}l\sigma A}) 
+\sum_{\mathbf{k}ll'\sigma}Q\epsilon_{\mathbf{k}12}
(\hat{c}^{\dagger}_{\mathbf{k}l\sigma A}\hat{c}_{\mathbf{k}l'\sigma
B}+\hat{c}^{\dagger}_{\mathbf{k}l\sigma
B}\hat{c}_{\mathbf{k}l'\sigma A})\\
&-\mu\sum_{\mathbf{k}l\sigma}(q^s_{\sigma
}\hat{n}_{\mathbf{k}l\sigma A}+q^s_{\bar{\sigma}}\hat{n}_{\mathbf{k}l\sigma
B})
+\frac{L}{2}\sum_{I_1,I_4} \bar{E}^A_{I_1,I_4}
\langle\hat{m}^A_{I_1,I_4}
\rangle_0 + \frac{L}{2}\sum_{I_1,I_4} \bar{E}^B_{I_1,I_4}
\langle\hat{m}^B_{I_1,I_4}
\rangle_0\;,
\label{eq:U_GA_recip_AF}
\end{split}
\end{equation}
where
\begin{equation}
 Q=q_{\uparrow}q_{\downarrow}-\bar{q}_{\uparrow}\bar{q}_{\downarrow}\;.
\end{equation}
It should be noted that the sums in (\ref{eq:U_GA_recip_AF}) are taken over all
$L/2$ independent $\mathbf{k}$ states.
As before, we apply the SGA method which leads to the effective
Hamiltonian with the statistical-consistency constraints of the form
\begin{equation}
\begin{split}
 \hat{K}_{\lambda}=\hat{K}_{GA}&-\lambda_{S}\bigg[\sum_{\mathbf{k}l\sigma}\frac{
1 } { 2 } \sigma(\hat{n}_{\mathbf{k}l\sigma A} - \hat{n}_{\mathbf{k}l\sigma
B})-2LS^z_{0|s} \bigg]\\
&-\lambda_+\bigg[\sum_{\mathbf{k}}(\hat{A}_{\mathbf{k}1A}+\hat{A}_{\mathbf{k}-1B
}) -L\Delta^0_+ + H.C.\bigg]\\
&-\lambda_-\bigg[\sum_{\mathbf{k}}(\hat{A}_{\mathbf{k}-1A}+\hat{A}_{\mathbf{k}
1B
}) -L\Delta^0_- + H.C. \bigg]\\
&-\lambda_n\bigg[\sum_{\mathbf{k}l\sigma}(q^s_{\sigma}\hat{n}_{\mathbf{k}l\sigma
A}+q^s_{\bar{\sigma}}\hat{n}_{ \mathbf{k}l\sigma B})-Ln_G \bigg]
\end{split}
\end{equation}
Introducing now the eight-component composite operator
\begin{equation}
%\begin{split}
 \mathbf{\hat{f}}^{\dagger}_{\mathbf{k}\sigma}\equiv(\hat{c}^{\dagger}_{\mathbf
{ k } 1
\sigma A} ,
 \hat{c}^{\dagger}_{\mathbf{k}2\sigma A}, \hat{c}^{\dagger}_{\mathbf{k}1\sigma
B},
 \hat{c}^{\dagger}_{\mathbf{k}2\sigma B}, \hat{c}_{-\mathbf{k}1
\sigma A} ,\hat{c}_{-\mathbf{k}2\sigma A}, \hat{c}_{-\mathbf{k}1\sigma B},
 \hat{c}_{-\mathbf{k}2\sigma B} )\;,
%\end{split}
\end{equation}
we can write down the effective Hamiltonian $\hat{K}_{\lambda}$ in the following
form
\begin{equation}
\begin{split}
 \hat{K}_{\lambda}&=\frac{1}{2}\sum_{\mathbf{k}\sigma}\mathbf{\hat{f}}^{\dagger}
_{\mathbf{k } \sigma }
\mathbf{\hat{M}}_{\mathbf{k}\sigma}\mathbf{\hat{f}}_{\mathbf{k}\sigma}
-(\mu+\lambda_n)(q^s_{\uparrow}+q^s_{\downarrow})L \\ 
&+2L\lambda_+\Delta_+^0+ 2L\lambda_-\Delta_-^0  + 2L\lambda_S S^z_{0|s} +
L\lambda_n n_G\\
&+\frac{L}{2}\sum_{I_1,I_4} \bar{E}^A_{I_1,I_4}
\langle\hat{m}^A_{I_1,I_4}
\rangle_0 + \frac{L}{2}\sum_{I_1,I_4} \bar{E}^B_{I_1,I_4}
\langle\hat{m}^B_{I_1,I_4}
\rangle_0\;,
\end{split}
\end{equation}
%\vspace{6 cm}
where the explicit form of the 8x8 matrix is
\begin{equation}
\mathbf{\hat{M}}_{\mathbf{k}\sigma}=\left(\begin{array}{cccccccc}
 \eta_{\sigma} & 0 & Q\epsilon_{\mathbf{k}} &
Q\epsilon_{\mathbf{k}12} & 0 & \lambda_{\sigma}^A & 0 & 0\\
0 & \eta_{\sigma} & Q\epsilon_{\mathbf{k}12} & Q\epsilon_{\mathbf{k}} &
-\lambda_{\sigma}^A & 0 & 0 & 0\\
Q\epsilon_{\mathbf{k}} & Q\epsilon_{\mathbf{k}12} &
\eta_{\sigma} & 0 & 0 & 0 & 0 & \lambda_{\sigma}^B\\
Q\epsilon_{\mathbf{k}12} & Q\epsilon_{\mathbf{k}} &
0 & \eta_{\sigma} & 0 & 0 & -\lambda_{\sigma}^B & 0\\
0 & -\lambda_{\sigma}^A & 0 & 0 & -\eta_{\sigma} & 0 & -Q\epsilon_{\mathbf{k}} &
-Q\epsilon_{\mathbf{k}12}\\
\lambda_{\sigma}^A & 0 & 0 & 0 & 0 & -\eta_{\sigma} & -Q\epsilon_{\mathbf{k}12}
& -Q\epsilon_{\mathbf{k}}\\
0 & 0 & 0 & -\lambda_{\sigma}^B & -Q\epsilon_{\mathbf{k}} &
-Q\epsilon_{\mathbf{k}12} &
-\eta_{\sigma} & 0\\
0 & 0 & \lambda_{\sigma}^B & 0 & -Q\epsilon_{\mathbf{k}12} &
-Q\epsilon_{\mathbf{k}} &
0 & -\eta_{\sigma}
\end{array} \right)\;,
\label{eq:matrix_H_AFSC}
\end{equation}
and
\begin{equation}
\begin{split}
 \lambda_{\uparrow}^A&=\lambda_{\downarrow}^B\equiv \lambda_+\;, \\
 \lambda_{\downarrow}^A&=\lambda_{\uparrow}^B\equiv \lambda_-\;, \\
 \eta_{\sigma}&=-\frac{1}{2}\sigma\lambda_S - q^s_{\sigma}(\mu+\lambda_n)\;.
\end{split}
\end{equation}
Diagonalization of (\ref{eq:matrix_H_AFSC}) leads to the quasi-particle
energies
$E_{\mathbf{k}l\sigma}$ ($l=1,2,3,...,8$). The corresponding grand potential
function
$F_{\lambda}$ per atomic site now has the form
\begin{equation}
\begin{split}
 F_{\lambda}&=-\frac{2}{L\beta}\sum_{\mathbf{k}l\sigma}\ln\big[1+e^{-\beta
E_{\mathbf{k}l\sigma}}
\big]-\mu(q^s_{\uparrow}+q^d_{\downarrow})\\
&+2\lambda_+\Delta_+^0+ 2\lambda_-\Delta_-^0  + 2\lambda_S S^z_{0|s} +
(\lambda_n+\mu)n_G\\
&+\frac{L}{2}\sum_{I_1,I_4} \bar{E}^A_{I_1,I_4}
\langle\hat{m}^A_{I_1,I_4}
\rangle_0 + \frac{L}{2}\sum_{I_1,I_4} \bar{E}^B_{I_1,I_4}
\langle\hat{m}^B_{I_1,I_4}
\rangle_0\;.
\end{split}
\end{equation}
As before, we minimize the $F_{\lambda}$ function to determine the values
of the mean fields, the variational parameters and the
Lagrange parameters. The necessary conditions for the minimum are again
expressed by
(\ref{eq:derivs}) and (\ref{eq:derivs_mu}). In the subsequent discussion we
consider also the pure antiferromagnetic
phase (AF), for which $S^z_{G|s}\neq0$ but $\Delta_{+}=\Delta_{-}\equiv0$. The
number of equations that need to be solved is different for different phases
considered in this work. In Table III we show how many equations are included
in (\ref{eq:derivs}) and (\ref{eq:derivs_mu}) for all phases discussed.

\begin{table}[h]
\centering
\caption{Number of equations that have to be solved in the case of all
considered here phases. To reduce the number of equations for particular phases
we have used certain symmetry relations regarding the mean field parameters,
the Lagrange multipliers, and the variational parameters.}
 \begin{tabular}{||c||c|c|c|c|c|c||}
  \hline
   phase       & A &  A1+FM & SC+AF & AF & NS & FM  \\ 
  \hline
   num. of Eq. & 16&  17    & 22    & 12 & 8  & 13  \\
  \hline
 \end{tabular}
 \label{table: values_tab}
\end{table}

%%%%%%%%%%%%%%%%%%%%%%%%%%%%%%%%%%%%%%%%%%%%%%%%%%%%%%%%%%%%%%%%%%%%%%%%%%%%%%%%
%%%%%%%%%%%%%%%%%%%%%%%%%%%%%%%%%%%%%%%%%%%%%%%%%%%%%%%%%%%%%%%%%%%%%%%%%%%%%%%%
\section{Results and discussion}

Equations (\ref{eq:derivs}) and (\ref{eq:derivs_mu}) 
have been solved numerically for all phases
by means of the hybrd1 subroutine from the MINPACK library, which performs a
modification of the Powell hybrid method. The maximal estimated error of the
procedure was set to $10^{-7}$. The derivatives in Eq. (\ref{eq:derivs}) and
(\ref{eq:derivs_mu}) were computed by using a 5-step stencil method
with the step equal to $x=10^{-4}$.

We concentrate now on the detailed numerical analysis of the
phase diagram and
the microscopic characterization of the stable phases.
Having in mind that for 3d orbitals $U'=U-2J$, one obtains the HF
condition for the
pairing to occur, $U<3J$ (see \cite{Zegrodnik2012}).
We discuss thus first and foremost the limit $U<3J$, as it allows for a
direct comparison of SGA with the HF solution. In this manner we can single
out explicitly
the role of correlations in stabilizing the relevant phases.
%%%%%%%%%%%%%%%%%%%%%%%%%%%%%%%%FIG1%%%%%%%%%%%%%%%%%%%%%%%%%%%%%%%%%%%%%%%%%%%%
\begin{figure}[b!]
\hfill
\epsfxsize=73mm 
\rotatebox{-90}{\epsfbox[15 11 423 880]{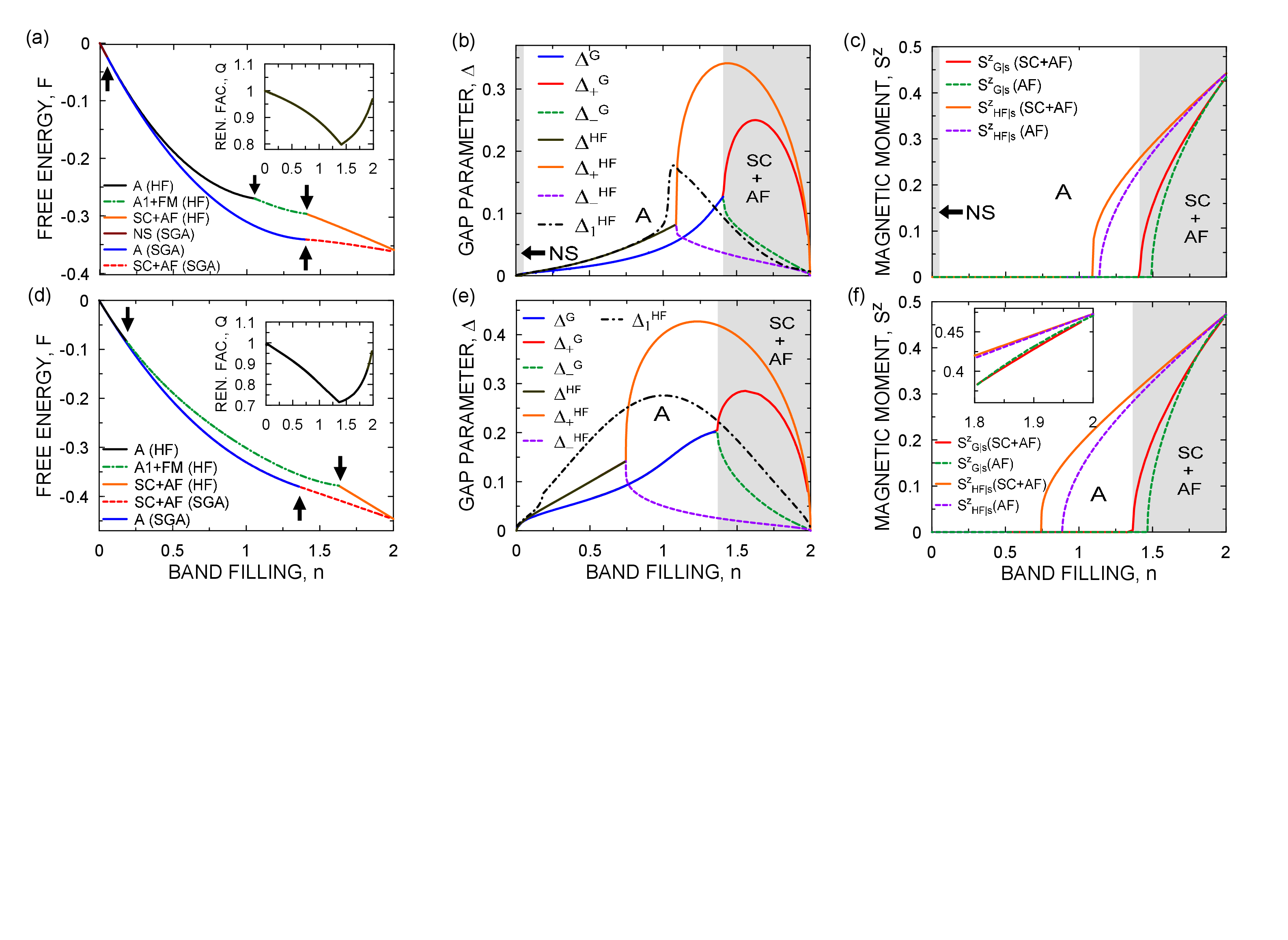}}
\caption{(Color online) Stable phases evolution vs. band filling. The
superconducting gap parameter, magnetic moment and
free energies as a function of band filling both for the HF and SGA, for
$J=0.299$: a, b, c and $J=0.4545$: d, e, f. The results
are for $\beta_h=0.0$. The shaded regions
represent the stability regions of the respective phases according to the SGA
calculations.
In
Figs. a and d we show only the free energies of stable phases. 
The arrows in a and d mark the transitions points
between
phases.}
\label{fig:1(a-f)}
\end{figure} 
One should note
that in the considered regime ($U<3J$) we have a model with 
intraatomic interorbital attractions
leading to spin-triplet pairs. 
As the main attractive force is of
intraatomic nature, we focus here on the local (s-wave) type of pairing
only. In other words, as we discuss the situation with no or small
hybridization, the intersite part of the pairing can be disregarded.%\newpage

The calculations have been carried out for $U'=U-2J$,
$U=2.2J$, $k_BT/W=10^{-4}$. This leaves us still with three independent
microscopic
parameters in
our model: $n_{G}$, $J$, and $\beta_h$. All the energies have been normalized to
the
bare band-width $W=8|t|$ (as we consider the square lattice 
 with nearest neighbor hopping). For
comparison, we also show the results calculated by means of the combined
HF-BCS$\equiv$HF approximation. This method is described in
detail in our previous paper for the same model as considered here. We can also
reproduce the HF results by using the Gutzwiller method described in this work 
 and setting
$\lambda_{I,I'}=\delta_{I,I'}$.

In Fig.~\ref{fig:1(a-f)} we display the free energy, superconducting
gaps, and magnetic moments for 
the two values $J=0.299$ and $J=0.4545$. 
As one can
see from the free-energy plots (Figs. \ref{fig:1(a-f)}a and 
\ref{fig:1(a-f)}d),
below some certain
value of band filling, the pure superconducting phase of type A is stable for
the SGA method. The increase of the number of electrons in the
system, enhances the gap in this
region (Figs. \ref{fig:1(a-f)}b and \ref{fig:1(a-f)}e). Above the critical band
filling $n_c$, the
staggered moment structure is created and a division into two gap parameters
($\Delta_+$ and $\Delta_-$) appears, as can be seen in Figs. \ref{fig:1(a-f)}b,
\ref{fig:1(a-f)}e, \ref{fig:1(a-f)}c, and \ref{fig:1(a-f)}f. In this regime the
SC+AF phase becomes stable.

%\newpage
When approaching half filling, both gaps gradually approach zero and for
$n=2$ we are left with a pure AF phase, which is of Slater insulating type
evolving towards the Mott-Hubbard insulating state with the
increasing U. As the
staggered
magnetic moment is rising (with the increase of
$n_G$), the renormalization factor is approaching unity (cf. Insets to
Fig.~\ref{fig:1(a-f)}a and \ref{fig:1(a-f)}d). This is a consequence
of the fact that for large values of $S^z_G$, the configurations with two
electrons of opposite spin, on the same orbital, are ruled out.
%%%%%%%%%%%%%%%%%%%%%%%%%%%%%%%%FIG2%%%%%%%%%%%%%%%%%%%%%%%%%%%%%%%%%%%%%%%%%%%%
\begin{figure}[h!]
\hfill
\epsfxsize=49mm 
\rotatebox{-90}{\epsfbox[16 12 249 656]{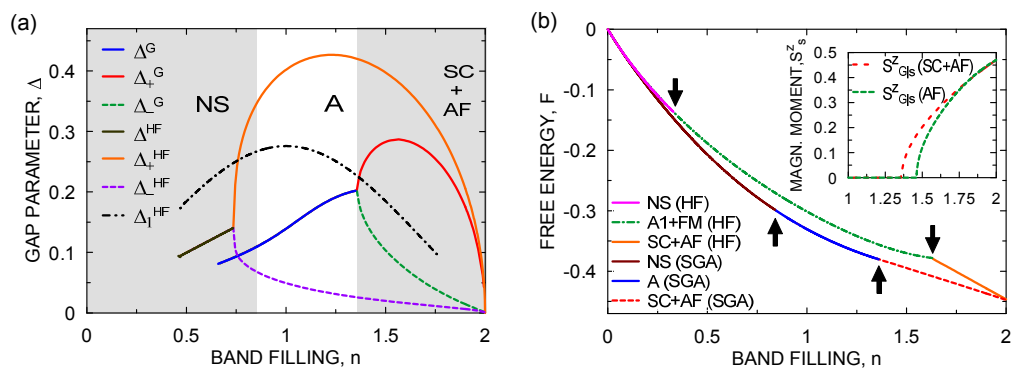}}
\caption{(Color online) The superconducting gaps (a) and the free energies (b)
as a function
of band filling for $J=0.4545$ and $\beta_h=0.1$. The shaded regions
represent stability of respective phases according to the SGA calculations. The
vertical arrows mark the phase borders.}
\label{fig:2(ab)}
\end{figure}

Comparing Figs. \ref{fig:1(a-f)}a, \ref{fig:1(a-f)}b, \ref{fig:1(a-f)}c with
Figs. \ref{fig:1(a-f)}d, \ref{fig:1(a-f)}e, \ref{fig:1(a-f)}f one
sees
that by increasing $J$ we make the value of $n_c$ smaller. However, the decrease
in $n_c$ is not as significant in SGA as it is in the
HF case. In general the results presented Figs. \ref{fig:1(a-f)}b,
\ref{fig:1(a-f)}c, \ref{fig:1(a-f)}e, \ref{fig:1(a-f)}f
look similar from the qualitative point of view for both 
methods.
For SGA, the onset of antiferromagnetically ordered phase
appears closer to half filling than for the HF method. Another
difference
between HF and SGA is that for the former the staggered moment in the
SC+AF phase is increased by the appearance of SC for the whole range of band
fillings,
whereas in
SGA calculations the staggered moment is slightly stronger in the AF phase
than in the SC+AF phase for a small region close to the half-filled situation
(inset
of Fig.~\ref{fig:1(a-f)}f).
%%%%%%%%%%%%%%%%%%%%%%%%%%%%%%%%FIG3%%%%%%%%%%%%%%%%%%%%%%%%%%%%%%%%%%%%%%%%%%%%
\begin{figure}[h!]
\hfill
\epsfxsize=45.5mm 
\rotatebox{-90}{\epsfbox[16 12 249 691]{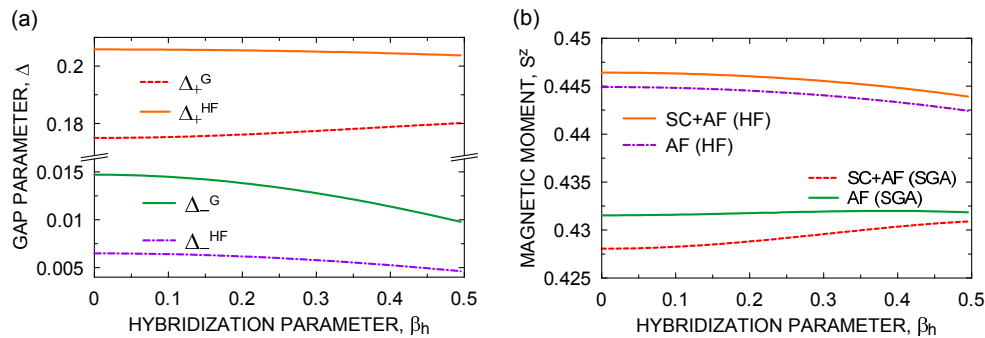}}
\caption{(Color online) The superconducting gaps and magnetic moment as a
function of the hybridization strength, $\beta_h$, for $n=1.9$, $J=0.4545$, for
the case of SC+AF and AF phases.}
\label{fig:3(ab)}
\end{figure}

Significant differences between HF and SGA can
be seen in Figs.
\ref{fig:1(a-f)}c and \ref{fig:1(a-f)}f. While changing the band filling from
0 to 2, in the case of SGA calculations we move consecutively through the
regions of stability of NS (for
$J=0.299$), A, SC+AF phases, and for $n=2$ we have pure antiferromagnetism. The
situation is different in the HF approximation, where in between the
regions
%%%%%%%%%%%%%%%%%%%%%%%%%%%%%%%%FIG4%%%%%%%%%%%%%%%%%%%%%%%%%%%%%%%%%%%%%%%%%%%%
\begin{figure}[b!]
\hfill
\epsfxsize=49mm 
\rotatebox{-90}{\epsfbox[16 12 249 656]{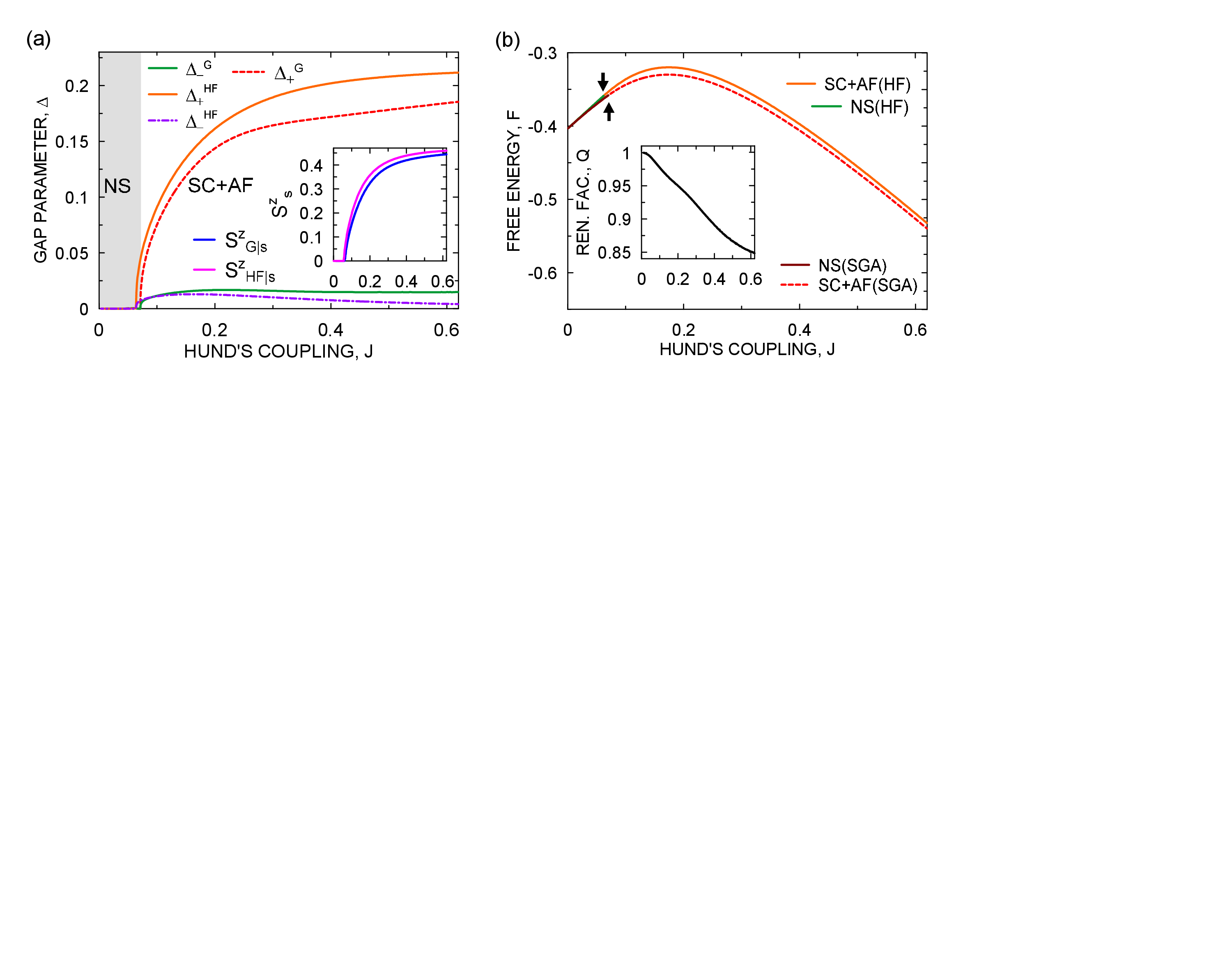}}
\caption{(Color online) The superconducting gaps in the SC+AF phase (a) and
free
energies of stable phases (b) as a function of Hund's coupling for $n=1.9$ and
$\beta_h=0.1$. The shaded region represent the stability of NS phase
according to the SGA results.}
\label{fig:4(ab)}
\end{figure}
of stability of A and SC+AF phase, we have also the stable A1+FM phase. It
should be also noted that the free energy calculated in SGA is lower
than the one for the HF situation, as one should expect, since the correlations
are accounted more accurately in the former method. It is also very interesting
that having the system with $U<3J$, no pure ferromagnetism appears in this
canonical model of itinerant magnetism.
%%%%%%%%%%%%%%%%%%%%%%%%%%%%%%%%FIG5%%%%%%%%%%%%%%%%%%%%%%%%%%%%%%%%%%%%%%%%%%%%
\begin{figure}[t!]
\hfill
\epsfxsize=86mm 
\rotatebox{-90}{\epsfbox[16 12 433 658]{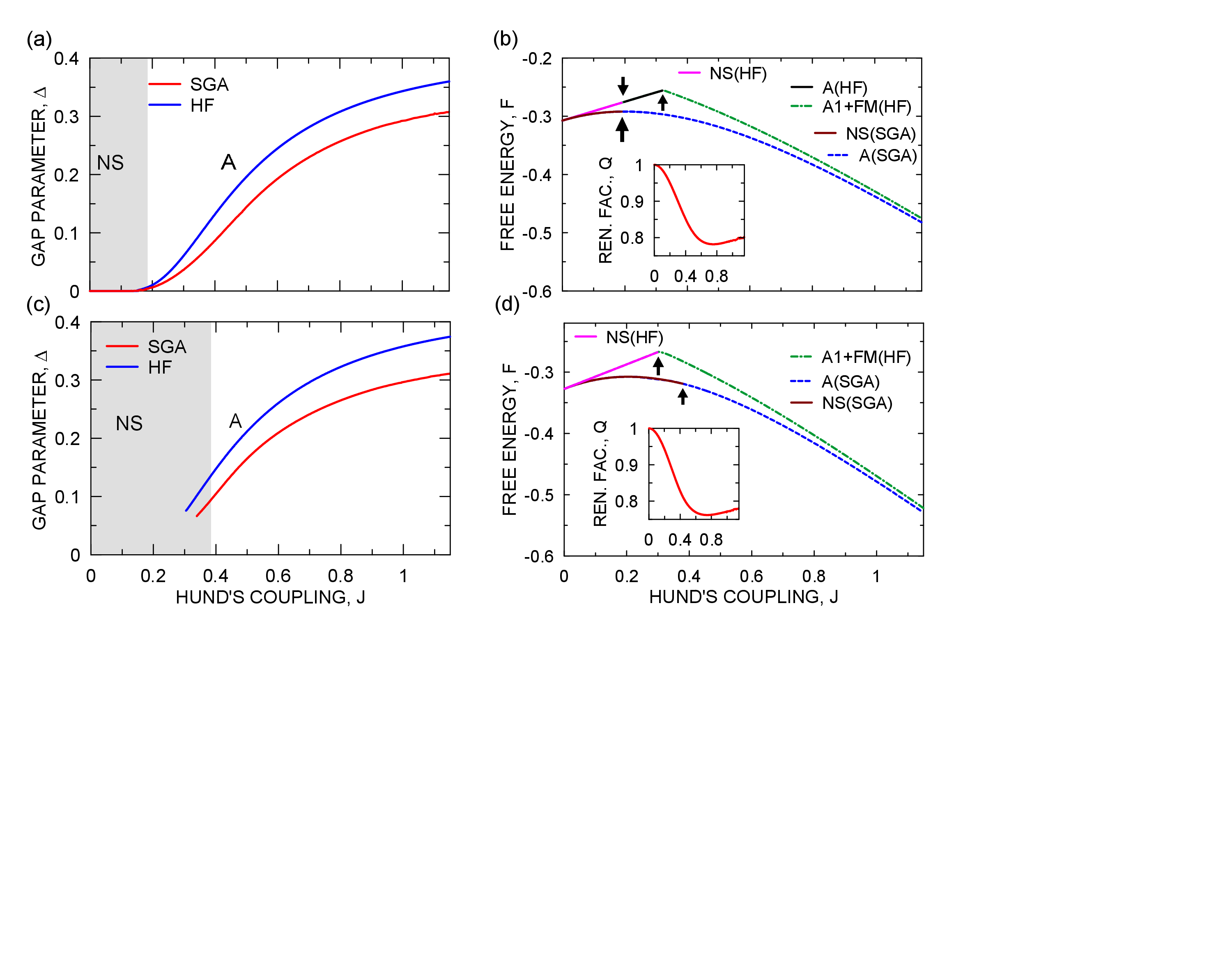}}
\caption{(Color online) The superconducting gaps in the A phase as a
function of the Hund's coupling for $n=1.0$ (a-for $\beta_h=0.0$ and c-for
$\beta_h=0.1$) and free energies of stable
phases corresponding to SGA and HF approximations (b-for $\beta_h=0.0$ and d-for
$\beta_h=0.1$). The shaded regions
represent the stability of the NS phase 
according to the SGA. The
vertical arrows mark the border points between respective phases. Insets:
Bandwidth renormalization factor for $\beta_h=0$ (upper) and $\beta_h=0.1$
(lower).}
\label{fig:5(ad)}
\end{figure}

In Fig.~\ref{fig:2(ab)}, we present the results for the case with nonzero
hybridization parameter, $\beta_h=0.1$. In this case there are no
superconducting solutions below some
certain value of the band filling 
(cf. Fig.~\ref{fig:2(ab)}a) and an extended
region of NS stability occurs. The
influence of the hybridization on the antiferromagnetically ordered phases is
weak, as can be seen more clearly in Fig.~\ref{fig:3(ab)}. The changes in
the superconducting gap
and the magnetic moment triggered by the hybridization, are
quite small even for larger values of $\beta_h$.

Next, we discuss the $J$ dependence of the superconducting gap, the free energy
and the
magnetic moment for selected values of band filling. As in the case
of $n$-dependences the gap parameters and the magnetic moments in both SGA
and HF approximation are qualitatively similar. In Fig.~\ref{fig:4(ab)} we can
see that for
$n=1.9$ even the
free-energy plots and regions of stability of certain phases are comparable for
both calculation schemes used. For the quarter-filled case
(cf. Fig.~\ref{fig:5(ad)}) the
A1+FM phase is stable above some value of
$J$, according to the HF results. However, this is not the case in the SGA
approximation, where the A phase has lower free energy even than the
saturated ferromagnetic phase coexisting with superconductivity. Comparing
Figs. \ref{fig:5(ad)}b and \ref{fig:5(ad)}d (as well as \ref{fig:1(a-f)}d and
\ref{fig:2(ab)}b) one sees
that the region of
stability of the A phase narrows down in
favor of the NS phase, due to the influence of hybridization. 

\begin{figure}[h]
\centering
\epsfxsize=75mm 
\epsfbox[14 558 351 782]{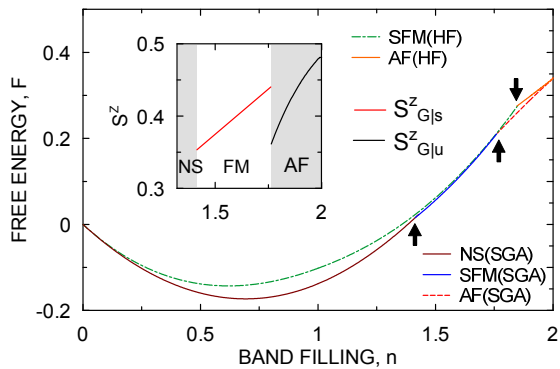}
\caption{(Color online) The free energies of the stable phases for SGA and HF
methods for $J=0.4$ and $U=1.6$. The shaded regions in the inset mark the
stability of certain phases according to the SGA approach. 
Note the appearance of
ferromagnetic phase for $U=4J$ (i.e., for $U>3J$) in the filling range
$1.45\div 1.75$, sandwiched in between paramagnetic and
antiferromagnetic phases.}
\label{fig:6}
\end{figure}

It is important to check whether the itinerant magnetic phases are stable in the
regime $U'>J$ ($U>3J$), i.e., when the superconductivity is absent in the HF
approximation. For this purpose, in Fig.~\ref{fig:6} we provide the band-filling
dependence of the
free
energy corresponding to stable phases for $U=4J$. Indeed, the
paramagnetic and the magnetically ordered phases
are
stable for both methods of calculations. Therefore, for $U>3J$ we recover
the magnetic phase diagram for this model, which was considered
originally only in
the
context of magnetism. The free energy of the saturated
ferromagnetic phase calculated by the SGA is very
close to the one obtained by the HF approach. This is again caused by the
circumstance
that in
the saturated state all of the spins are parallel 
and the double occupancies on
the same orbital are absent. In this situation, the intra-orbital Coulomb
interaction is automatically switched off. It would be interesting to determine
the stability of the
coexistent phases in this regime ($U'>J$). Work along this line is in progress. 
%%%%%%%%%%%%%%%%%%%%%%%%%%%%%%%%FIG6%%%%%%%%%%%%%%%%%%%%%%%%%%%%%%%%%%%%%%%%%%%%

\begin{table}[h]
\centering
\caption{Exemplary values of the order parameters, the chemical
potential, the free energy, and the band renormalization factors
corresponding to the considered phases, for two different sets of
values of the microscopic parameters $n$ and $J$. The
underlined values correspond the stable phases. The numerical accuracy is on
the level of the last digit specified.}
 \begin{tabular}{cccc}
  \hline\hline
             &       &  $n=1.0$ & $n=1.9$ \\ 
   parameter & phase &  $J=0.299$ & $J=0.299$ \\
  \hline
  $\Delta$   &   A   & \underline{0.0450027} & 0.1701940 \\
  $\Delta$   & A1+FM & 0.0426749 & 0.1307664 \\
  $\Delta_+$ & SC+AF & -         & \underline{0.1638992}  \\
  $\Delta_-$ & SC+AF & -         & \underline{0.0161868}  \\
  \hline
  $S^z_u$    & A1+FM  & 0.000317  & 0.1092674  \\
  $S^z_s$    & SC+AF  & -         & \underline{0.3902738}  \\
  $S^z_s$    &  AF    & -         & 0.3885899  \\
  \hline
  $\mu$     &    A    &\underline{-0.1382377} & 0.16078601 \\
  $\mu$     &   NS    &-0.1377649 & 0.1875964  \\
  $\mu$     & A1+FM   &-0.1379700 & 0.18222514 \\
  $\mu$     &  SC+AF  &-          & \underline{-0.0421144} \\
  $\mu$     &   AF    &-          & -0.0893963 \\
  \hline
  $F$       &    A    &\underline{-0.3106091} & -0.3118381 \\
  $F$       &   NS    &-0.3105145 & -0.2992516 \\
  $F$       & A1+FM   &-0.3105586 & -0.3020254 \\
  $F$       &  SC+AF  &-          & \underline{-0.3576542} \\
  $F$       &   AF    &-          & -0.3509731 \\
  \hline
  $Q_{\uparrow}$  & A1+FM & 0.8845776 & 0.6751619 \\
  $Q_{\downarrow}$& A1+FM & 0.8839251 & 0.6282452 \\
  $Q$             &  A    & \underline{0.8845136} & 0.6736373 \\
  $Q$             &  NS   & 0.8840340 & 0.6421089 \\
  $Q$             & SC+AF &-          & \underline{0.9211224} \\
  $Q$             &  AF   &-          & 0.9293567 \\
  \hline\hline
 \end{tabular}
 \label{table: values_tab}
\end{table}

\section{Conclusions and outlook}
\subsection{Conclusions}
The principal purpose of this paper was to formulate a many-particle method which allows to investigate
the spin-triplet real-space pairing in correlated system with an orbital
degeneracy. To this end, we have carried out a detailed
analysis using the statistically-consistent
Gutzwiller approximation (SGA) for the 
two-band degenerate Hubbard model with
the spin-triplet superconductivity and itinerant
magnetism included, both treated on equal footing.
The results were compared with those coming from the Hartree-Fock approximation
amended with the Bardeen-Cooper-Schrieffer (BCS) approach.
The obtained Hund's
coupling and band filling dependences of the
magnetic moment and the superconducting gap
parameters are
often similar from the qualitative point of view with those evaluated by means
of the
HF approximation. However, the stability regions of the
considered
phases are significantly different for the two applied methods. In SGA, the
stable
coexisting superconducting-ferromagnetic phase is absent
while it appears in the HF approximation in a certain range of $J$ and
$n$ values. Furthermore, the coexistence of the paired state
with antiferromagnetism
appears much closer to the half-filled situation in SGA than in HF
approximation. For $n=2$ the superconductivity disappears and only the pure
antiferromagnetism 
survives; this state can be termed a correlated Slater-insulator state, which
evolves gradually into the Mott-Hubbard insulating state with increasing $U>1$
and $S^z_{G|s}\rightarrow 1/2$. 

The influence of hybridization for both approximations is similar.
With an increase of the $\beta_h$ parameter, the region of stability of the
superconducting type-A phase narrows down in favor of the NS state. On the other
hand, the
antiferromagnetic phase is not affected in any significant manner by 
an increase of 
$\beta_h$.

The band renormalization factors approach unity as the
interaction
constants $J$, $U$ and $U'$ tend to zero, what represents an additional
test of our numerical results correctness. Generally, in the low-coupling limit
our present results reduce to those obtained in HF approximation analysed by us
in \cite{Zegrodnik2012}, as it should be.
%The diagramatic approach for the multiband
%Gutzwiller approximation in general uses a large number of variational
%parameters which can create problems during numerical calculations especially
%for
%more than two bands or for anisotropic $3d$ band of $e_g$ type. However, a
%reasonable reduction of the number of nonzero,
%off-diagonal variational parameters makes the problem tractable without any
%significant loss of precision.

It
is important to emphasize that for both approaches the phase diagrams have been obtained for $U<3J$, i.e.
for relatively low value of the Hubbard interaction U, or equivalently, for a
relatively high value of
the Hund's rule exchange integral. 
A complete analysis of the present model would require studying the stability of
the spin-triplet superconductivity and its coexistence with magnetic ordering
in the
complementary regime $U>3J$, where the magnetism is favored against
superconductivity. This regime has been the subject 
in a number of
earlier papers
\cite{Bunemann1998,Kunes2010,Deng2009}, as then both
the
intraorbital, as well as the interorbital interaction is
repulsive, and lead in a
natural manner to magnetic ordering.

\subsection{Outlook: Extension and application to real systems}
In connection with the remarks provided above, we would like to characterize
briefly the possibility of extending the present model (\ref{eq:H_start}) to the
uranium systems in which superconductivity and ferromagnetism coexist in an
unambiguous manner \cite{Jedrak2011_2}. First of all, the magnetic moment in
those systems, particularly in UGe$_2$ and URhGe, is quite large, 
with an associated  molecular field in the megagauss 
range, which 
most probably rules
out any spin-singlet character of pairing (note that the Curie
temperature ($T_C$) to the superconducting transition temperature ($T_S$) ratio
reaches in the uranium compounds the value $T_C/T_S\sim 10^2$). In spite of
those circumstances, our solution does not provide any extended regime (for
the studied parameter range) for the ferromagnetism-spin-triplet
superconductivity
coexistence. Instead, in a wide range of band fillings, the coexisting
SC+AF phase is stable (cf. Fig.~\ref{fig:1(a-f)}c and \ref{fig:1(a-f)}f), as
well as the pure
spin-triplet superconducting phase of type A (the equal-spin-paired phase).
The pure A
phase seems to be realized in Sr$_2$RuO$_4$, though then a detailed
three-orbital
structure of the order parameter seems to be relevant \cite{Annet2009}. 
A
direct application of our SGA scheme to a realistic three-band system is more involved, as the number of parameters
to minimize would lead to a
computing time-consuming procedure, but still possible to
tackle.

The extension of the present model to the uranium system such as UGe$_2$ would
require
considering orbitally degenerate and correlated $5f^2-5f^3$
quasi-atomic states
due to U and hybridized with the uncorrelated conduction band states. This means
that we must
have minimally a three-orbital system with two partially occupied $5f$
quasi-atomic
states (so the Hund's rule becomes operative) and at least one extra conduction
band. Such situation may lead to a partial
Mott-localization phenomenon, i.e., to a spontaneous decomposition of $5f^n$
($n>1$)
configuration of electrons into the localized and the
itinerant
parts \cite{Zwicknagl2006}. In
such a situation, it is possible that the localized electrons are the source of
ferromagnetism, whereas the itinerant particles are paired \cite{Visser2010}.
This is not the type of coexistent phase we have in mind here, since in the
model considered by us all the
system electrons are indistinguishable in the quantum-mechanical sense. 

These considerations lead to the conclusion
that one would require minimally a periodic Anderson model with
degenerate
$5f^n$ states,
to mimic the uranium-based ferromagnetic superconductors.
This variant of the multiple-band
model is also very useful in the discussion of heavy-fermion
compounds.
Moreover, in the systems represented by this model, the coexistence of
antiferromagnetism and superconductivity has
been shown to appear in both experiment \cite{Knebel2011} and
theory \cite{Howczakpress}. More specifically for the systems
UPt$_3$ and UNi$_2$Al$_3$ the coexistence
of the spin-triplet pairing and the antiferromagnetism has already been
suggested to appear \cite{Morel,Anderson1994}, although not elaborated
in any detail. One specific feature should be mentioned. Namely, in the
situation when we have antiferromagnetic superconductor, then there is also a
strong theoretical indication that there is a spin-triplet component even for
the pure spin-singlet mechanism
of pairing \cite{Kaczmarczyk2011,Howczakpress}.
The spin-triplet important admixture results simply from a decomposition of the
system into two sublattices with staggered magnetic moment. These and related
features must be taken into quantitative analysis before any realistic
consideration of concrete systems is carried out. 
%In this respect, our paper
%represents the first attempt albeit in the model situation, incorporating the
%exchange interaction-induced spin-triplet pairing for correlated electrons into
%the canonical model of itinerant magnetism.

In connection with the whole discussion, it is intriguing to ask if a
symmetric model
system of the type exemplified by Hamiltonian (\ref{eq:H_start}) could be
experimentally realized in the optical lattice. Some
model systems (e.g., the Hubbard model system) have been experimentally
achieved in this manner \cite{Bloch2008}.

In relation to the spin-triplet real-space pairing induced by the
Hund's rule, one should also mention the spin
fluctuations (SF) as a possible mechanism of spin-triplet pairing in both
magnetic \cite{Bloch2008} and liquid $^3$He systems \cite{Fay1980}. Within the
present approach the spin fluctuations should be treated as quantum
fluctuations around the present self-consistently renormalized mean field
state \cite{Anderson1977}. The real-space and the spin-fluctuation contributions
may
become of comparable magnitude in the close vicinity of the quantum
critical point, where the ferro- or antiferro- states disappear under e.g.
pressure. This is, however, a completely separate topic of studies.

%%%%%%%%%%%%%%%%%%%%%%%%%%%%%%%%%%%%%%%%%%%%%%%%%%%%%%%%%%%%%%%%%%%%%%%%%%%%%%%%
\section{Acknowledgements}
Disscusions with Jakub J\k{e}drak and Jan Kaczmarczyk are gratefully
acknowledged. M.Z. has been partly supported by the EU Human Capital Operation
Program, Polish
Project No. POKL.04.0101-00-434/08-00. J.S. acknowledges the financial support
from the Foundation for Polish Science (FNP) within project TEAM. The grant
MAESTRO from the National Science Center (NCN) was helpful for the PL-DE
cooperation within the present project on a unified approach to magnetism and
superconductivity in correlated fermion systems.

%%%%%%%%%%%%%%%%%%%%%%%%%%%%%%%%%%%%%%%%%%%%%%%%%%%%%%%%%%%%%%%%%%%%%%%%%%%%%%%%

\end{document}